\documentclass[prd,preprint,superscriptaddress,nofootinbib]{revtex4}

\usepackage{amsmath,amssymb,array,calc,epsfig}
\usepackage{graphicx}

\newlength{\intwidth}

\begin{document}

\preprint{UWO-TH-07/09}

\title{Consistency Conditions On The S-Matrix Of Massless Particles}

\author{Paolo Benincasa}
\email{pbeninca@uwo.ca}

\affiliation{Department of Applied Mathematics, University of
Western Ontario, London, Ontario N6A 5B7, Canada}

\author{Freddy Cachazo}
\email{fcachazo@perimeterinstitute.ca}

\affiliation{Perimeter Institute for Theoretical Physics, Waterloo,
Ontario N2J 2W9, Canada}

\begin{abstract}

We introduce a set of consistency conditions on the S-matrix of
theories of massless particles of arbitrary spin in four-dimensional
Minkowski space-time. We find that in most cases the constraints,
derived from the conditions, can only be satisfied if the S-matrix
is trivial. Our conditions apply to theories where four-particle
scattering amplitudes can be obtained from three-particle ones via a
recent technique called BCFW construction. We call theories in this
class constructible. We propose a program for performing a
systematic search of constructible theories that can have
non-trivial S-matrices. As illustrations, we provide simple proofs
of already known facts like the impossibility of spin $s > 2$
non-trivial S-matrices, the impossibility of several spin 2
interacting particles and the uniqueness of a theory with spin 2 and
spin 3/2 particles.

\end{abstract}

\maketitle

\section{Introduction}

The power of the constraints that Lorentz invariance imposes on the
S-matrix of four dimensional theories has been well known at least
since the work of Weinberg \cite{Weinberg:1964ev, Weinberg:1964ew}.
Impressive results like the impossibility of long-range forces
mediated by massless particles with spin $>2$, charge conservation
in interactions mediated by a massless spin $1$ particle, or the
universality of the coupling to a massless spin $2$ particle are
examples beautifully obtained by simply using the pole structure of
the S-matrix governing soft limits in combination with Lorentz
invariance \cite{Weinberg:1964ew, Weinberg:1965rz}.

Weinberg's argument does not rule out the possibility of non-trivial
Lagrangians describing self-interacting massless particles of higher
spins. It rules out the possibility of those fields producing
macroscopic effects. Actually, the theory of massless particles of
higher spins has been an active research area for many years (see
reviews \cite{Sorokin:2004ie, Bouatta:2004kk} and reference therein,
also see \cite{Saitou:2006ca, Bastianelli:2007pv} for alternative
approaches). Lagrangians for free theories have been well understood
while interactions have been a stumbling block. Recent progress
shows that in spaces with negative cosmological constant it is
possible to construct consistent Lagrangian theories but no similar
result exists for flat space-time
\cite{Fotopoulos:2006ci,Buchbinder:2006eq}. Despite the difficulties
of constructing an interactive Lagrangian, several attempts have
been made in studying the consistency of specific couplings among
higher spin particles. For example, cubic interactions have been
studied in \cite{Berends:1984wp, Fradkin:1987ks, Fradkin:1986qy,
Deser:1990bk}. Also, very powerful techniques for constructing
interaction vertices systematically have been developed using
BRST-BV cohomological methods 
\cite{Henneaux:1997bm,Bekaert:2006us,Fotopoulos:2007nm} and references
therein.

In this paper we introduce a technique for finding theories of
massless particles that can have non-trivial S-matrices within a
special set of theories we call constructible. The starting point is
always assuming a Poincar{\'e} covariant theory where the S-matrix
transformation is derived from that of one-particle states which are
irreducible representations of the Poincar{\'e} group. There will
also be implicit assumptions of locality and parity invariance.

The next step is to show that for complex momenta, on-shell
three-particle S-matrices of massless particles of any spin can be
uniquely determined. As is well known, on-shell three-particle
amplitudes vanish in Minkowski space. That this need not be the case
for amplitudes in signatures different from Minkowski or for complex
momenta was explained by Witten in \cite{Witten:2003nn}.

We consider theories for which four-particle tree-level S-matrix
elements can be completely determined by three-particle ones. These
theories are called {\it constructible}. This is done by introducing
a one parameter family of complex deformations of the amplitudes and
using its pole structure to reconstruct it. The physical
singularities are on-shell intermediate particles connecting
physical on-shell three-particle amplitudes. This procedure is known
as the BCFW construction \cite{Britto:2004ap, Britto:2005fq}. One
can also introduce the terminology {\it fully constructible} if this
procedure can be extended to all $n$-particle amplitudes. Examples of
fully constructible theories are Yang-Mills \cite{Britto:2005fq} and
General Relativity \cite{Bedford:2005yy, Cachazo:2005ca,
Benincasa:2007qj} (the fact that cubic couplings could play
a key role in Yang-Mills theory and General Relativity was already understood
in \cite{Boulware:1974sr, Deser:1969wk}).

The main observation is that by using the BCFW deformation, the
four-particle amplitude is obtained by summing over only a certain
set of channels, say the ${\sf s}$- and the ${\sf u}$- channels.
However, if the theory under consideration exists, then the answer
should also contain the information about the ${\sf t}$- channel. In
particular, one could construct the four-particle amplitude using a
different BCFW deformation that sums only over the ${\sf t}$- and
the ${\sf u}$- channel.

Choosing different deformations for constructing {\it the same}
four-particle amplitude and requiring the two answers to agree is
what we call the four-particle test. This simple consistency
condition turns out to be a powerful constraint that is very
difficult to satisfy.

It is important to mention that the constraints are only valid for
constructible theories. Luckily, the set of constructible theories
is large and we find many interesting results. We also discuss some
strategies for circumventing this limitation.

As illustrations of the simplicity and power of the four-particle
test we present several examples. The first is a general analysis of
theories of a single spin $s$ particle. We find that if $s>0$ all
theories must have a trivial S-matrix except for $s=2$ which passes
the test. As a second example we allow for several particles of the
same spin. We find that, again in the range $s>0$, the only theories
that can have a nontrivial S-matrix are those of spin $1$ with
completely antisymmetric three-particle coupling constants which
satisfy the Jacobi identity and spin 2 particles with completely
symmetric three-particle coupling constants which define a
commutative and associative algebra. We also study the possible
theories of particles of spin $s$, without self-couplings and with
$s>1$, that can couple non-trivially to a spin $2$ particle. In this
case, we find that only $s=3/2$ passes the test. Moreover, all
couplings in the theory must be related to that of the three-spin-2
particle amplitude. Such a theory is linearized ${\cal N} =1$
supergravity.

The paper is organized as follows. In section II, we review the
construction of the S-matrix and of scattering amplitudes for
massless particles. In section III, we discuss how three-particle
amplitudes are non-zero and uniquely determined up the choice of the
values of the coupling constant. In section IV, we apply the BCFW
construction to show how, for certain theories, four-particle
amplitudes can be computed from three-particle ones. A theory for
which this is possible is called {\it constructible}. We then
introduce the four-particle test. In section V, we discuss
sufficient conditions for a theory to be constructible. In section
VI, we give examples of the use of the four-particle test. In
section VII, we conclude with a discussion of possible future
directions including how to relax the constructibility constraint.
Finally, in the appendix we illustrate one of the methods to relax
the constructibility condition.

\section{Preliminaries}

\subsection{S-Matrix}

In this section we define the S-matrix and scattering amplitudes. We
do this in order to set up the notation. Properties of the S-matrix,
which we exploit in this paper like factorization, have been well
understood since at least the time of the S-matrix program
\cite{Olive:1964, Chew:1966, Olive:1967}.

Recall that physically, one is interested in the probability for, say,
two asymptotic states to scatter and to produce $n-2$ asymptotic
states. Any such probability can be computed from the matrix
elements of momentum eigenstates
\begin{equation}
\left._{\rm out}\langle p_1\ldots p_{n-2} | p_a p_b\rangle_{\rm in}
\right. = \langle p_1\ldots p_{n-2} |S| p_a p_b\rangle
\end{equation}
where $S$ is a unitary operator. As usual, it is convenient to write
$S= {\mathbb I}+i T$ with
\begin{equation}
\langle p_1\ldots p_{n-2} |iT| p_a p_b\rangle =
\delta^{(4)}\left(p_a+p_b-\sum_{i=1}^{n-2}p_i\right)
M(p_a,p_b\rightarrow \{p_1,p_2,\ldots, p_{n-2}\}).
\end{equation}
$M(p_a,p_b\rightarrow \{p_1,p_2,\ldots, p_{n-2}\})$ is called the
scattering amplitude (see for example chapter 4 in \cite{Peskin:1995ev}).

Assuming crossing symmetry one can write $p_a=-p_{n-1}$ and
$p_b=-p_{n}$ and introduce a scattering amplitude where all
particles are outgoing. Different processes are then obtained by
analytic continuation of
\begin{equation}
M_n = M_n(p_1,p_2,\ldots, p_{n-1},p_{n}).
\end{equation}

$M_n$ is our main object of study. Our goal is to determine when
$M_n$ can be non-zero. Up to now we have exhibited only the
dependence on momenta of external particles. However, if they have
spin $s>0$ one also has to specify their free wave functions or
polarization tensors. We postpone the discussion of the explicit
form of polarization tensors until section V.

\subsection{Massless Particles Of Spin $s$}

It turns out that all the information needed to describe the
physical information of an on-shell massless spin $s$ particle is
contained in a pair of spinors $\{ \lambda_a,\tilde\lambda_{\dot
a}\}$, left- and right-handed respectively, and the helicity of the
particle~\cite{Berends:1981rb,DeCausmaecker:1981bg,Kleiss:1985yh,Witten:2003nn}.
Recall that in a Poincar{\'e} invariant theory, irreducible massless
representations are classified by their helicity which can be $h =
\pm s$ with $s$ any integer or half-integer known as the spin of the
particle.

The spinors $\{ \lambda_a,\tilde\lambda_{\dot a}\}$ transform in the
representations $(1/2,0)$ and $(0,1/2)$ of the universal cover of
the Lorentz group, $SL(2,{\mathbb C})$, respectively. Invariant
tensors are $\epsilon^{ab}$, $\epsilon^{\dot a\dot b}$ and
$(\sigma^\mu)_{a\dot a}$ where $\sigma^\mu = ({\mathbb
I},\vec{\sigma})$. The most basic Lorentz invariants, from which any
other is made of, can be constructed as follows:
\begin{equation}
\lambda_a\lambda'_b\epsilon^{ab} \equiv \langle \lambda,
\lambda'\rangle , \qquad \tilde\lambda_{\dot a}\tilde\lambda'_{\dot
b}\epsilon^{\dot a\dot b} \equiv [ \lambda, \lambda' ].
\end{equation}
Finally, using the third invariant tensor we can define the momentum
of the particle by $p^\mu = \lambda^a (\sigma^\mu)_{a\dot
a}\tilde\lambda^{\dot a}$, where indices are raised using the first
two tensors. A simple consequence of this is that the scalar product
of two vectors, $p^\mu$ and $q^\mu$ is given by $2p\cdot q = \langle
\lambda^p ,\lambda^q\rangle [\tilde\lambda^p,\tilde\lambda^q]$.

\section{Three Particle Amplitudes: A Uniqueness Result}

In this section we prove that three-particle amplitudes of massless
particles of any spin can be uniquely determined.

The statement that on-shell scattering amplitudes of three massless
particles can be non-zero might be somewhat surprising. However, as
shown by Witten \cite{Witten:2003nn}, three-particle amplitudes are
naturally non-zero if we choose to work with the complexified
Lorentz group $SL(2,{\mathbb C})\times SL(2,{\mathbb C})$, where
$(1/2,0)$ and $(0,1/2)$ are completely independent representations
and hence momenta are not longer real. In other words, if
$\tilde\lambda_{\dot a} \neq \pm \bar\lambda_a$ then $p^\mu$ is
complex.

Let us then consider a three-particle amplitude
$M_3(\{\lambda^{(i)},\tilde\lambda^{(i)},h_i\})$ where the spinors
of each particle, $\lambda^{(i)}$ and $\tilde\lambda^{(i)}$, are
independent vectors in ${\mathbb C}^2$.

Momentum conservation $(p_1+p_2+p_3)_{a\dot a}=0$ and the on-shell
conditions, $p_i^2=0$, imply that $p_i\cdot p_j = 0$ for any $i$ and
$j$. Therefore we have the following set of equations
\begin{equation}
\label{nosi} \langle 1,2\rangle [1,2] = 0, \qquad \langle 2,3\rangle
[2,3] = 0, \qquad \langle 3,1\rangle [3,1] = 0.
\end{equation}

Clearly, if $[1,2]=0$ and $[2,3]=0$ then $[3,1]$ must be zero. The
reason is that the spinors live in a two dimensional vector space
and if $\tilde\lambda^{(1)}$ and $\tilde\lambda^{(3)}$ are
proportional to $\tilde\lambda^{(2)}$ then they must also be
proportional.

This means that the non-trivial solutions to (\ref{nosi}) are either
$\langle 1,2\rangle = \langle 2,3\rangle= \langle 3,1\rangle = 0$ or
$[1,2]=[2,3]=[3,1]=0$.

Take for example $[1,2]=[2,3]=[3,1]=0$ and set
$\tilde\lambda^{(2)}_{\dot a} = \alpha_2 \tilde\lambda^{(1)}_{\dot
a}$ and $\tilde\lambda^{(3)}_{\dot a} = \alpha_3
\tilde\lambda^{(1)}_{\dot a}$. Then momentum conservation implies
that $\lambda^{(1)}_a + \alpha_2 \lambda^{(2)}_a+\alpha_3
\lambda^{(3)}_a = 0$ which is easily seen to be satisfied if
$\alpha_2=-\langle 1,3\rangle/\langle 2,3\rangle$ and
$\alpha_3=-\langle 1,2\rangle/\langle 3,2\rangle$.

The conclusion of this discussion is that three-particle amplitudes,
$M_3(\{\lambda^{(i)},\tilde\lambda^{(i)},h_i\})$, which by Lorentz
invariance are only restricted to be a generic function of $\langle
i,j\rangle$ and $[i,j]$ turn out to split into a ``holomorphic" and
an ``anti-holomorphic" part\footnote{Using ``holomorphic" and
``anti-holomorphic" is an abuse of terminology since
$\tilde\lambda_{\dot a} \neq \pm \bar\lambda_a$. We hope this will
not cause any confusion.}. More explicitly
\begin{equation}
M_3 = M^H_3(\langle 1,2\rangle,\langle 2,3\rangle,\langle
3,1\rangle) + M^A_3([1,2],[2,3],[3,1]).
\end{equation}

It is important to mention that we are considering the full
three-particle amplitude and not just the tree-level one. Therefore
$M^H_3$ and $M^A_3$ are not restricted to be rational
functions\footnote{We thank L. Freidel for discussions on this
point.}. In other words, we have purposefully avoided to talk about
perturbation theory. We will be forced to do so later in section V
but we believe that this discussion can be part of a more general
analysis.

\subsection{Helicity Constraint and Uniqueness}

One of our basic assumptions about the S-matrix is that the
Poincar{\'e} group acts on the scattering amplitudes as it acts on
individual one-particle states. This in particular means that the
helicity operator must act as
\begin{equation}
\label{helioper} \left( \lambda_i^a\frac{\partial}{\partial
\lambda_i^a} - \tilde\lambda_i^a\frac{\partial}{\partial
\tilde\lambda_i^a}\right) M_3(1^{h_1},2^{h_2},3^{h_3}) = -2h_i
M_3(1^{h_1},2^{h_2},3^{h_3}).
\end{equation}
Equivalently,
\begin{equation}
\label{helima} \left( \lambda_i^a\frac{\partial}{\partial
\lambda_i^a} + 2h_i\right) M^H_3(\langle 1,2\rangle,\langle
2,3\rangle,\langle 3,1\rangle) = 0
\end{equation}
on the holomorphic one and as
\begin{equation}
\label{antihelima} \left( \tilde\lambda_i^a\frac{\partial}{\partial
\tilde\lambda_i^a} -2h_i\right) M^A_3([1,2],[2,3],[3,1]) = 0
\end{equation}
on the anti-holomorphic one.

It is not difficult to show that if $d_1=h_1-h_2-h_3$,
$d_2=h_2-h_3-h_1$ and $d_3=h_3-h_1-h_2$, then
\begin{equation}
F = \langle 1,2\rangle^{d_{3}}\langle 2,3\rangle^{d_{1}}\langle
3,1\rangle^{d_{2}}, \qquad G = [1,2]^{-d_{3}}[ 2,3]^{-d_{1}}[
3,1]^{-d_{2}}
\end{equation}
are particular solutions of the equations (\ref{helima}) and
(\ref{antihelima}) respectively.

Therefore, $M^H_3/F$ and $M^A_3/G$ must be ``scalar" functions,
i.e., they have zero helicity.

Let $x_1$ be either $\langle 2,3\rangle$ or $[2,3]$ depending on
whether we are working with the holomorphic or the antiholomorphic
pieces. Also let $x_2$ be either $\langle 3,1\rangle$ or $[3,1]$ and
$x_3$ be either $\langle 1,2\rangle$ or $[1,2]$. Finally, let ${\cal
M}$ be either $M^H_3/F$ or $M^A_3/G$. Then we find that
\begin{equation}
x_i\frac{\partial {\cal M}(x_1,x_2,x_3)}{\partial x_i} = 0
\end{equation}
for $i=1,2,3$. Therefore, up to solutions with delta function
support which we discard based on analyticity, the only solution for
${\cal M}$ is a constant. Let such a constant be denoted by
$\kappa_H$ or $\kappa_A$ respectively.

We then find that the exact three-particle amplitude must be
\begin{equation}
M_3(\{\lambda^{(i)},\tilde\lambda^{(i)},h_i\}) = \kappa_H \langle
1,2\rangle^{d_{3}}\langle 2,3\rangle^{d_{1}}\langle
3,1\rangle^{d_{2}}+ \kappa_A [1,2]^{-d_{3}}[ 2,3]^{-d_{1}}[
3,1]^{-d_{2}}.
\end{equation}

Finally, we have to impose that $M_3$ has the correct physical
behavior in the limit of real momenta. In other words, we must
require that $M_3$ goes to zero when both $\langle i,j\rangle$ and
$[i,j]$ are taken to zero\footnote{Taking to zero $\langle
i,j\rangle$ means that $\lambda^{(i)}$ and $\lambda^{(j)}$ are
proportional vectors. Therefore, all factors $\langle i,j\rangle$
can be taken to be proportional to the same small number $\epsilon$
which is then taken to zero.}. Simple inspection shows that if
$d_1+d_2+d_3$, which is equal to $-h_1-h_2-h_3$, is positive then we
must set $\kappa_A =0$ in order to avoid an infinity while if
$-h_1-h_2-h_3$ is negative then $\kappa_H$ must be zero. The case
when $h_1+h_2+h_3 = 0$ is more subtle since both pieces are allowed.
In this paper we restrict our study to $h_1+h_2+h_3 \neq 0$ and
leave the case $h_1+h_2+h_3 = 0$ for future work.

\subsection{Examples}

Let us consider few examples, which will appear in the next
sections, as illustrations of the uniqueness of three-particle
amplitudes.

Consider a theory of several particles of a given integer spin s.
Since all particles have the same spin we can replace $h =\pm s$ by
the corresponding sign. Let us use the middle letters of the
alphabet to denote the particle type.

There are only four helicity configurations:
\begin{equation}
\label{exione} M_3(1^-_m,2^-_r,3^+_s) = \kappa_{mrs} \left(
\frac{\langle 1,2\rangle^3}{\langle 2,3\rangle\langle
3,1\rangle}\right)^s, \qquad M_3(1^+_m,2^+_r,3^-_s) = \kappa_{mrs}
\left( \frac{[ 1,2 ]^3}{ [2,3] [3,1]}\right)^s
\end{equation}
and
\begin{equation}
\label{exitwo} M_3(1^-_m,2^-_r,3^-_s) = \kappa'_{mrs} \left( \langle
1,2\rangle\langle 2,3\rangle\langle 3,1\rangle\right)^s, \qquad
M_3(1^+_m,2^+_r,3^+_s) = \kappa'_{mrs} \left( [ 1,2 ][2,3]
[3,1]\right)^s.
\end{equation}
The subscripts on the coupling constants $\kappa$ and $\kappa'$ mean
that they can depend on the particle type\footnote{Note that here we
have implicitly assumed parity invariance by equating the couplings
of conjugate amplitudes.}. We will use the amplitudes in (\ref{exione}) 
in section VI.

A simple but important observation is that if the spin is odd then
the coupling constant must be completely antisymmetric in
its indices. This is because due to crossing symmetry the
amplitude must be invariant under the exchange of labels.

This leads to our first result, a theory of less than three massless
particles of odd spin must have a trivial three-particle S-matrix.
Under the conditions of constructibility, this can be extended to
higher-particle sectors of the S-matrix and even to the full
S-matrix.

\section{The Four-Particle Test And Constructible Theories}

In this section we introduce what we call the four-particle test.
Consider a four-particle amplitude $M_4$. Under the assumption that
one-particle states are stable in the theory, $M_4$ must have poles
and multiple branch cuts emanating from them at locations where
either ${\sf s}=(p_1+p_2)^2$, ${\sf t}=(p_2+p_3)^2$ or ${\sf
u}=(p_3+p_1)^2$ vanish\footnote{We have introduced the notation
${\sf s}$ for the center of mass energy in order to avoid confusion
with the spin $s$ of the particles.}.

We choose to consider only the pole structure. Branch cuts will
certainly lead to very interesting constraints but we leave this for
future work. Restricting to the pole structure corresponds
to working at tree-level in field theory.

As we will see, under certain conditions, one can construct physical
on-shell tree-level four-particle amplitudes as the product of two
on-shell three-particle amplitudes (evaluated at complex momenta
constructed out of the real momenta of the four external particles)
times a Feynman propagator. In general this can be done in at least
two ways. Roughly speaking, these correspond to summing over the
${\sf s}$-channel and ${\sf u}$-channel or summing over the ${\sf
t}$-channel and ${\sf u}$-channel. A necessary condition for the
theory to exists is that the two four-particle amplitudes
constructed this way give the same answer. This is what we call the
four-particle test. It might be surprising at first that a sum over
the ${\sf s}$- and ${\sf u}$-channels contains information about the
${\sf t}$-channel but as we will see this is a natural consequence
of the BCFW construction which we now review.

\subsection{Review Of The BCFW Construction And Constructible Theories}

The key ingredient for the four-particle test is the BCFW
construction \cite{Britto:2005fq}. The construction can be applied to
$n$-particle amplitudes, but for the purpose of this paper we only need
four-particle amplitudes.

We want to study $M_4(\{\lambda^{(i)}_a, \tilde\lambda^{(i)}_{\dot
a}, h_i\})$. Recall that momenta constructed from the spinors of
each particle are required to satisfy momentum conservation, i.e.,
$(p_1+p_2+p_3+p_4)^\mu = 0$.

Choose two particles, one of positive and one of negative
helicity\footnote{Here we do not consider amplitudes with all equal
helicities.}, say $i^{+s_i}$ and $j^{-s_j}$, where $s_i$ and $s_j$
are the corresponding spins, and perform the following deformation
\begin{equation}
\label{bcfw} \lambda^{(i)}(z) = \lambda^{(i)} + z\lambda^{(j)},
\qquad \tilde\lambda^{(j)}(z) = \tilde\lambda^{(j)} -
z\tilde\lambda^{(i)}.
\end{equation}
All other spinors remain the same.

The deformation parameter $z$ is a complex variable. It is easy to
check that this deformation preserves the on-shell conditions, i.e.,
$p_k(z)^2 = 0$ for any $k$ and momentum conservation since
$p_i(z)+p_j(z) = p_i+p_j$.

The main observation is that the scattering amplitude is a rational
function of $z$ which we denote by $M_4(z)$. This fact follows from
$M_4(1^{h_1},\ldots , 4^{h_4})$ being, at tree-level, a rational
function of spinor products. Being a rational function of $z$,
$M_4(z)$ can be determined if complete knowledge of its poles,
residues and behavior at infinity is found.

\bigskip

{\bf Definition:} We call a theory {\it constructible} if $M_4(z)$
vanishes at $z=\infty$. As we will see this means that $M_4(z)$ can
only be computed from $M_3$ and hence the name.

\bigskip

In the next section we study sufficient conditions for a theory to
be constructible. The proof of constructibility relies very strongly
on the fact that on-shell amplitudes should only produce the two
physical helicity states of a massless particle\footnote{This in
turn is simply a consequence of imposing Lorentz invariance
\cite{Weinberg:1964ew}.}. In this section we assume that the theory 
under consideration is constructible.

Any rational function that vanishes at infinity can be written as a
sum over its poles with the appropriate residues. In the case at
hand, $M_4(z)$ can only have poles of the form
\begin{equation}
\frac{1}{(p_i(z)+p_k)^2} = \frac{1}{\langle
\lambda^{(i)}(z),\lambda^{(k)}\rangle [i,k]} = \frac{1}{(\langle i,
k\rangle + z\langle j, k\rangle )[i,k]}
\end{equation}
where $k$ has to be different from $i$ and $j$.

As mentioned at the beginning of this section, $M_4(z)$ can be
constructed as a sum over only two of the three channels. The reason
is the following. For definiteness let us set $i=1$ and $j=2$, then
the only propagators that can be $z$-dependent are
$1/(p_1(z)+p_4)^2$ and $1/(p_1(z)+p_3)^2$. By construction
$1/(p_1+p_2)^2$ is $z$-independent.

The rational function $M_4(z)$ can thus be written as
\begin{equation}
M_4^{(1,2)}(z) = \frac{c_{\sf t}}{z-z_{\sf t}} + \frac{c_{\sf
u}}{z-z_{\sf u}}
\end{equation}
where $z_{\sf t}$ is such that ${\sf t}=(p_1(z)+p_4)^2$ vanishes,
i.e., $z_{\sf t} = -\langle 1, 4\rangle/\langle 2, 4\rangle$ while
$z_{\sf u}$ is where ${\sf u}=(p_1(z)+p_3)^2$ vanishes, i.e.,
$z_{\sf u} = -\langle 1, 3\rangle/\langle 2, 3\rangle$. Note that we
have added the superscript $(1,2)$ to $M_4(z)$ to indicate that it
was obtained by deforming particles $1$ and $2$.

\begin{figure}
\[M_{4}^{(1,2)}\:=\:
 \sum_{h}\!\!\!\raisebox{2.23cm}{\scalebox{.85}{{\includegraphics*[116pt,735pt][281pt,592pt]{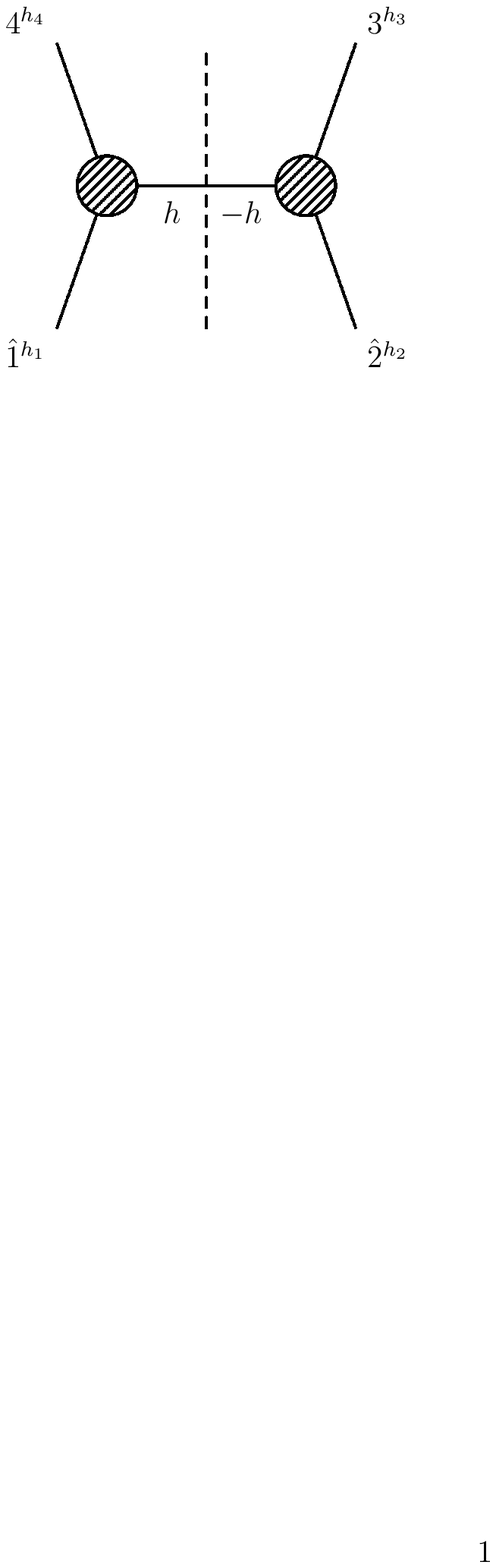}}}}\!\!\frac{1}{P_{14}^2}\;+\;
 \sum_{h}\!\!\!\raisebox{2.23cm}{\scalebox{.85}{{\includegraphics*[116pt,735pt][281pt,592pt]{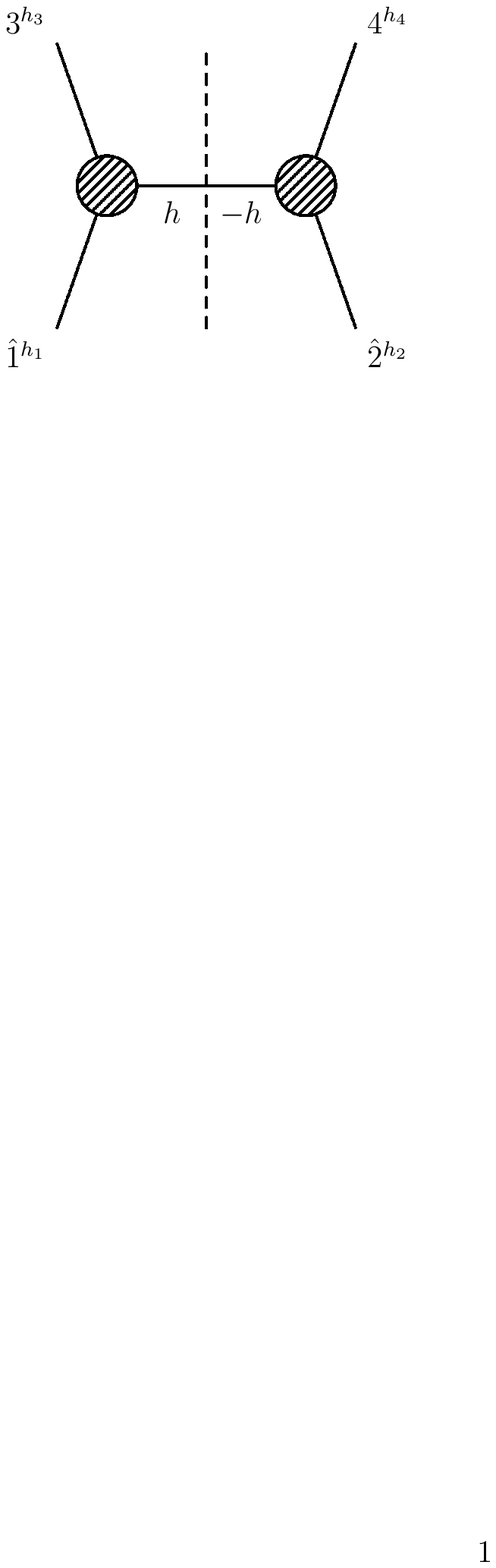}}}}\!\!\frac{1}{P_{13}^2}
\]
\vspace{1cm}
\caption{Factorization of a four-particle amplitude into two on-shell three-particle amplitudes.  
         In constructible theories, four-particle amplitudes are given by a sum over simple poles
         of the 1-parameter family of amplitudes $M_{4}(z)$ times the corresponding residues. At
         the location of the poles the internal propagators go on-shell and the residues are the
         product of two on-shell three-particle amplitudes.\label{fig:Factorization}}
\end{figure}

Finally, we need to compute the residues. Close to the location of
one of the poles, $M_4(z)$ factorizes as the product of two on-shell
three-particle amplitudes. Note that each of the three-particle
amplitudes is on-shell since the intermediate particle is also
on-shell. See figure~\ref{fig:Factorization} for a schematic representation. Therefore, we
find that
\begin{equation}
\begin{split}
M_4^{(1,2)}(z)  = & \sum_h M_3(p_1^{h_1}(z_{\sf t}), p_4^{h_4},
-P_{1,4}^h(z_{\sf t}))\frac{1}{P_{1,4}^2(z)}M_3(p_2^{h_2}(z_{\sf t}),
p_3^{h_3}, P_{1,4}^{-h}(z_{\sf t})) + \\ & \sum_h
M_3(p_1^{h_1}(z_{\sf u}), p_3^{h_3}, -P_{1,3}^h(z_{\sf
u}))\frac{1}{P_{1,3}^2(z)}M_3(p_2^{h_2}(z_{\sf u}), p_4^{h_4},
P_{1,3}^{-h}(z_{\sf u})).
\end{split}
\end{equation}
where the sum over $h$ runs over all possible helicities in the
theory under consideration and also over particle types if there is
more than one.

The scattering amplitude we are after is simply obtained by setting
$z=0$, i.e, $M_4(\{\lambda^{(i)},\tilde\lambda^{(i)}, h_i\}) =
M_4^{(1,2)}(0)$.

Recall that we assumed $h_1 = s_1$ and $h_2 = -s_2$. Let us further
assume that $h_4 = -s_4$. Therefore we could repeat the whole
procedure but this time deforming particles $1$ and $4$. In this way
we should find that $M_4(\{\lambda^{(i)},\tilde\lambda^{(i)}, h_i\})
= M_4^{(1,4)}(0)$.

We have finally arrived at the consistency condition we call the
four-particle test. One has to require that
\begin{equation}
\label{testy} M_4^{(1,2)}(0) = M_4^{(1,4)}(0).
\end{equation}

%
%
%
%

As we will see in examples, this is a very strong condition that
very few constructible theories satisfy non-trivially. In other
words, most constructible theories satisfy (\ref{testy}) only if all
three-particle couplings are set to zero and hence four-particle
amplitudes vanish. If the theory is fully constructible, this
implies that the whole S-matrix is trivial.

\subsection{Simple Examples}

We illustrate the use of the four-particle test by first working out
the general form of $M_4^{(1,2)}(0)$ and $M_4^{(1,4)}(0)$ for a
theory containing only integer spin particles\footnote{Including
half-integer spins is straightforward and we give an example in
section VI.}. We then specialize to the case of a theory containing
a single particle of integer spin $s$. It turns out that the theory 
is constructible only when $s>0$. For $s>0$, we explicitly find the
condition on $s$ for the theory to pass the four-particle test.

\subsubsection{General Formulas For Integer Spins}

Consider first $M_4^{(1,2)}(0)$. In order to keep the notation
simple we will denote $\langle\lambda^{(1)}(z), \bullet\rangle$ by
$\langle \hat 1, \bullet\rangle$ and so on. The precise value of $z$
depends on the deformation and channel being considered.
\begin{equation}
\label{geko}
\begin{split}
M_4^{(1,2)}(0) = & \sum_h \left(  \kappa^H_{(1+h_1+h_4+h)}
\langle \hat 1,4\rangle^{h-h_1-h_4}\langle 4,{\hat P}_{1,4}\rangle^{h_1-h_4-h}
\langle {\hat P}_{1,4} , \hat 1 \rangle^{h_4-h-h_1} + \right. \\
& \left. \kappa^A_{(1-h_1-h_4-h)}[1,4]^{h_1+h_4-h}[4,{\hat
P}_{1,4}]^{h_4+h-h_1}[{\hat P}_{1,4},1]^{h+h_1-h_4}\right)
\times\frac{1}{P_{1,4}^2}\times
\\ & \left( \kappa^H_{(1+h_2+h_3-h)} \langle 3,2\rangle^{-h-h_3-h_2}\langle
2,{\hat P}_{1,4}\rangle^{h_3-h_2+h}\langle {\hat P}_{1,4},3\rangle^{h_2-h_3+h} \right. + \\
 & \left. \kappa^A_{(1-h_2-h_3+h)}[3,\hat 2]^{h+h_3+h_2}[\hat 2,{\hat
P}_{1,4}]^{-h_3+h_2-h}[{\hat P}_{1,4},3]^{-h_2+h_3-h}\right) + \\
& \sum_h (4\leftrightarrow 3).
\end{split}
\end{equation}
Here the subscripts on the three-particle couplings denote the
dimension of the coupling. The range of values of the helicity of
the internal particle depends on the details of the specific theory
under consideration. Even though (\ref{geko}) is completely general
we choose to exclude theories where $h$ can take values such that
$h+h_1+h_2 =0$ or $-h+h_2+h_3 =0$. The main reason is that formulas
will simplify under this assumption.

Note also that we have kept the two pieces of all three-particle
amplitudes entering in (\ref{geko}). However, recall that we should
set either the holomorphic or the anti-holomorphic coupling to zero.
As we will now see this condition is very important for the
consistency of (\ref{geko}).

Let us solve the condition $P_{1,4}(z)^2 =0$. As mentioned above
this leads to $z_t=-\langle 1,4\rangle/\langle 2,4\rangle$. Since
$P_{1,4}(z_t)$, which we denoted by ${\hat P}_{1,4}$, is a null
vector, it must be possible to find spinors $\lambda^{(\hat P)}$ and
$\tilde\lambda^{(\hat P)}$ such that ${\hat P}_{1,4}^\mu =
\lambda^{(\hat P)a}(\sigma^\mu)_{a\dot a}\tilde\lambda^{(\hat P)\dot
a}$. Clearly, given ${\hat P}_{1,4}$ it is not possible to uniquely
determine the spinors since any pair of spinors $\{ t\lambda^{(\hat
P)}, t^{-1}\tilde\lambda^{(\hat P)}\}$ gives rise to the same ${\hat
P}_{1,4}$. This ambiguity drops out of (\ref{geko}) as we will see.

After some algebra we find that
\begin{equation}
P_{1,4}(z_t) = {\hat P}_{1,4} =
\frac{[1,4]}{[1,3]}\lambda^{(4)}\tilde\lambda^{(3)}.
\end{equation}
Therefore we can choose
\begin{equation}
\lambda^{(\hat P)} = \alpha \lambda_4, \quad \tilde\lambda^{(\hat
P)} = \beta \tilde\lambda_3, \quad {\rm with} \quad \alpha\beta
=\frac{[1,4]}{[1,3]}.
\end{equation}
Moreover, it is also easy to get
\begin{equation}
\hat\lambda_1 = \frac{\langle 2,1\rangle}{\langle
2,4\rangle}\lambda_4, \quad \hat{\tilde\lambda}_2 =
\frac{[1,2]}{[1,3]}\tilde\lambda_3.
\end{equation}

Using the explicit form of all the spinors one can check that the
three-particle amplitude with coupling constant
$\kappa^H_{(1+h_1+h_4+h)}$ in (\ref{geko}) possesses a factor of the
form $\langle 4,4\rangle = 0$ to the power $-h_1-h_4-h$. From our
discussion in section III, if $-h_1-h_4-h$ is less than zero then
the coupling $\kappa^H_{(1+h_1+h_4+h)}=0$. In this way a possible
infinity is avoided. Therefore we get a contribution from the term
with coupling $\kappa^A_{(1-h_1-h_4-h)}$ whenever $h>-(h_1+h_4)$.

Now, if $-h_1-h_4-h$ is positive then $\kappa^H_{(1+h_1+h_4+h)}$
need not vanish but the factor multiplying it vanishes. In this case
$\kappa^A_{(1-h_1-h_4-h)}$ must be zero and we find no
contributions.This means that the only non-zero contributions to the 
sum over $h$ can only come from the region where $h>-(h_1+h_4)$.

Turning to the other three-particle amplitude, we find that the
piece with coupling $\kappa^A_{(1-h_2-h_3+h)}$ has a factor
$[3,3]=0$ to the power $-h+h_2+h_3$. A similar analysis shows that
the only nonzero contributions come from regions where
$h>(h_2+h_3)$.

Putting the two conditions together we find that the first term
gives a non-zero contribution only when $h> {\rm
max}(-(h_1+h_4),(h_2+h_3))$.

Simplifying we find
\begin{equation}
\label{hasi}
\begin{split}
M_4^{(1,2)}(0) = & \sum_{h > {\rm max}(-(h_1+h_4),(h_2+h_3))} \left(
\kappa^A_{1-h_1-h_4-h}\kappa^H_{1+h_2+h_3-h}
\frac{(-P_{3,4}^2)^h}{P_{1,4}^2}\left(\frac{[1,4][3,4]}{[1,3]}\right)^{h_4}
\right. \\
& \left.
\left(\frac{[1,3][1,4]}{[3,4]}\right)^{h_1}\left(\frac{\langle
3,4\rangle}{\langle 2,3\rangle\langle
2,4\rangle}\right)^{h_2}\left(\frac{\langle 2,4\rangle}{\langle
2,3\rangle\langle 3,4\rangle}\right)^{h_3}\right) + \sum_{h > {\rm
max}(-(h_1+h_3),(h_2+h_4))}\!\!\!\!\!\!\!\!(4\leftrightarrow 3).
\end{split}
\end{equation}

Finally, it is easy to obtain $M_4^{(1,4)}(0)$ from (\ref{hasi}) by
simply exchanging the labels $2$ and $4$.

Next we will write down all formulas explicitly for the case when
$|h_i|=s$ for all $i$.

\subsubsection{Theories Of A Single Spin s Particle}

Consider now the case $h_1 = s$, $h_2 = -s$, $h_3 = s$ and $h_4 =
-s$. We also assume that the theory under consideration has a single
particle of spin s. This restriction is again for simplicity. If one
decided to allow for more internal particles then the different
terms would have to satisfy the four-particle test independently
since the dimensions of the coupling constants would be
different\footnote{There might be cases where the dimensions might
agree by accident. Such cases might actually lead to new interesting
theories. We briefly elaborate in section VII but we leave the
general analysis for future work.}.

Using (\ref{hasi}) we find that the first sum contributing to
$M^{(1,2)}_4(0)$ allows only for $h=s$ while the second one allows
for $h=-s$ and $h=s$. Using momentum conservation\footnote{One can
easily show that momentum conservation for four particles implies
that $\langle a,b\rangle/\langle a,c\rangle = - [d,c]/[d,b]$ for any
choice of $\{a,b,c,d\}$.} to simplify the expressions we find
\begin{equation}
\label{pola}
\begin{split}
M_4^{(1,2)}(0) = & \kappa^A_{1-s}\kappa^H_{1-s}\left(\frac{\langle
2,4\rangle^3[1,3]}{\langle 1,2\rangle\langle 3,4\rangle}\right)^s
\frac{1}{\langle 1,4\rangle [1,4]} +
\kappa^A_{1-s}\kappa^H_{1-s}\left(\frac{[1,3]^3\langle
4,2\rangle}{[4,3][1,2]}\right)^s \frac{1}{\langle 1,3\rangle [1,3]}
+ \\
& \kappa^A_{1-3s}\kappa^H_{1-3s}([1,3]\langle
4,2\rangle)^{2s}\frac{\left(-P_{3,4}^2\right)^s}{P_{1,3}^2}.
\end{split}
\end{equation}

We would like to set all couplings with the same dimension to the
same value. In other words, we define $\kappa =\kappa^A_{1-s} =
\kappa^H_{1-s}$. We also choose to study the case $\kappa'
=\kappa^A_{1-3s} = \kappa^H_{1-3s} = 0$. It turns out that if we had
chosen $\kappa=0$ and $\kappa'$ non-zero the resulting theories
would not have been constructible. In section VII we explore
strategies for relaxing this condition.

As mentioned above we can write $M^{(1,4)}(0)$ by simply exchanging
the labels 2 and 4. We then find
\begin{equation}
\label{seca}
\begin{split}
M_4^{(1,2)}(0) = & \kappa^2 \left(\frac{\langle
2,4\rangle^3[1,3]}{\langle 1,2\rangle\langle 3,4\rangle}\right)^s
\frac{1}{\langle 1,4\rangle [1,4]} +
\kappa^2\left(\frac{[1,3]^3\langle 4,2\rangle}{[4,3][1,2]}\right)^s
\frac{1}{\langle 1,3\rangle [1,3]},
 \\
M_4^{(1,4)}(0) = & \kappa^2 \left(\frac{\langle
4,2\rangle^3[1,3]}{\langle 1,4\rangle\langle 3,2\rangle}\right)^s
\frac{1}{\langle 1,2\rangle [1,2]} +
\kappa^2\left(\frac{[1,3]^3\langle 2,4\rangle}{[2,3][1,4]}\right)^s
\frac{1}{\langle 1,3\rangle [1,3]}.
\end{split}
\end{equation}

Both amplitudes can be further simplified to
\begin{equation}
M_4^{(1,2)}(0) =  -(-1)^s \kappa^2 \frac{\left( [1,3]\langle
2,4\rangle\right)^{2s}}{{\sf s t u}} \times {\sf s}^{2-s}, \quad
M_4^{(1,4)}(0) =  -(-1)^s \kappa^2 \frac{\left( [1,3]\langle
2,4\rangle\right)^{2s}}{{\sf s t u}} \times {\sf t}^{2-s}.
\end{equation}

Finally, the four-particle test requires $M_4^{(1,2)}(0)
=M_4^{(1,4)}(0)$ or equivalently $M_4^{(1,2)}(0)/M_4^{(1,4)}(0) =
1$. The latter gives the condition $({\sf s}/{\sf t})^{2-s} =1$
which can only be satisfied for generic choices of kinematical
invariants if $s=2$. If $s\neq 2$ the four-particle test
$M_4^{(1,2)}(0) =M_4^{(1,4)}(0)$ then requires $\kappa =0$ and hence
a trivial S-matrix.

\section{Conditions For Constructibility}

The example in the previous section showed that the only theory of a
single massless spin s particle that passes the four-particle test
is that with $s=2$. This theory turns out to be linearized General
Relativity. For $s=1$, the result is also familiar: a single photon
should be free. However, if $s=0$ one knows that a single scalar can
have a non-trivial S-matrix. The reason we did not find $s=0$ as a
possible solution in the previous example is that precisely for
$s=0$ the four-particle amplitude is not constructible. Therefore
our calculation was valid only for $s>0$.

In this section we study the criteria for constructibility in more
detail. Unfortunately, we do not know a way of carrying out this
discussion without first assuming the existence of a Lagrangian. The
conditions for constructibility will therefore be given in terms of
conditions on the interaction vertices of a Lagrangian. We will also
assume that it is possible to perform a perturbative expansion using
Feynman diagrams. The starting point of all theories we consider is
a canonical kinetic term (free Lagrangian) which for $s=0,1,2$ is
very well known and for $s>2$ can be found for example in
\cite{Fronsdal:1978rb, Buchbinder:1998qv, Sorokin:2004ie}.

The first ingredient is the polarization tensors of massless
particles of spin $s$. Polarization tensors of particles of integer
spin $s$ can be expressed in terms of polarization vectors of spin
$1$ particles as follows:
\begin{equation}
\epsilon^+_{a_1\dot a_1, \ldots ,a_s\dot a_s} =
\prod_{i=1}^s\epsilon_{a_i\dot a_i}^+ \, , \qquad
\epsilon^-_{a_1\dot a_1, \ldots ,a_s\dot a_s} =
\prod_{i=1}^s\epsilon_{a_i\dot a_i}^- .
\end{equation}
For half-integer spin $s+1/2$ they are
\begin{equation}
\epsilon^+_{a_1\dot a_1, \ldots ,a_s\dot a_s,\dot b} =
\tilde\lambda_{\dot b}\prod_{i=1}^s\epsilon_{a_i\dot a_i}^+ \, ,
\qquad \epsilon^-_{a_1\dot a_1, \ldots ,a_s\dot a_s, b} =
\lambda_b\prod_{i=1}^s\epsilon_{a_i\dot a_i}^- ,
\end{equation}
and where polarization vectors of spin $1$ particles are given by
\begin{equation}
\epsilon^+_{a\dot a} = \frac{\mu_a\tilde\lambda_{\dot a}}{\langle
\mu ,\lambda \rangle}, \qquad \epsilon^-_{a\dot a}
=\frac{\lambda_a\tilde\mu_{\dot a}}{[\tilde\lambda , \tilde\mu]}
\end{equation}
with $\mu_a$ and $\tilde\mu_{\dot a}$ arbitrary reference spinors.

This explains how all the physical data of a massless particle can
be recovered from $\lambda, \tilde\lambda$ and $h$. A comment is in
order here. The presence of arbitrary reference spinors means that
polarization tensors cannot be uniquely fixed once $\{ \lambda,
\tilde\lambda ,h\}$ is given. If a different reference spinor is
chosen, say, $\mu'$ for $\epsilon^+_{a\dot a}$ then
\begin{equation}
\label{asom} \epsilon^+_{a\dot a}(\mu') = \epsilon^+_{a\dot a}(\mu)
+ \omega \lambda_a\tilde\lambda_{\dot a}
\end{equation}
where
$$\omega =\frac{ \langle\mu',\mu\rangle}{\langle\mu',\lambda\rangle\langle\lambda,
\mu\rangle}.$$

If the particle has helicity $h=1$ then it is easy to recognize
(\ref{asom}) as a gauge transformation and the amplitude must be
invariant.

However, one does not have to invoke gauge invariance or assume any
new principle. As shown by Weinberg in \cite{Weinberg:1964ew} for any
spin $s$, the only way to guarantee the correct Poincar{\'e} transformations
of the S-matrix of massless particles is by imposing invariance under
(\ref{asom}). In that sense, there is no assumption in this
section that has not already been made in section II. In other words, 
Poincar{\'e} symmetry requires that $M_n$ gives the
same answer independently of the choice of reference spinor $\mu$.

\subsection{Behavior at Infinity}

If a theory comes from a Lagrangian then the three-particle
amplitudes derived in section III can be computed as the product of
three polarization tensors times a three-particle vertex that
contains some power of momenta which we denote by $L_3$. Simple
dimensional arguments indicate that if all particles have integer
spin then $L_3 = |h_1+h_2+h_3|$. Let us denote the power of momenta
in the four-particle vertex by $L_4$.

We are interested in the behavior of $M_4$, constructed using
Feynman diagrams, under the deformation of $\lambda^{(1)}$ and
$\tilde\lambda^{(2)}$ defined in (\ref{bcfw}) as $z$ is taken to
infinity.

\begin{figure}
\[
  \raisebox{2.15cm}{\scalebox{.85}{{\includegraphics*[121pt,733pt][277pt,598pt]{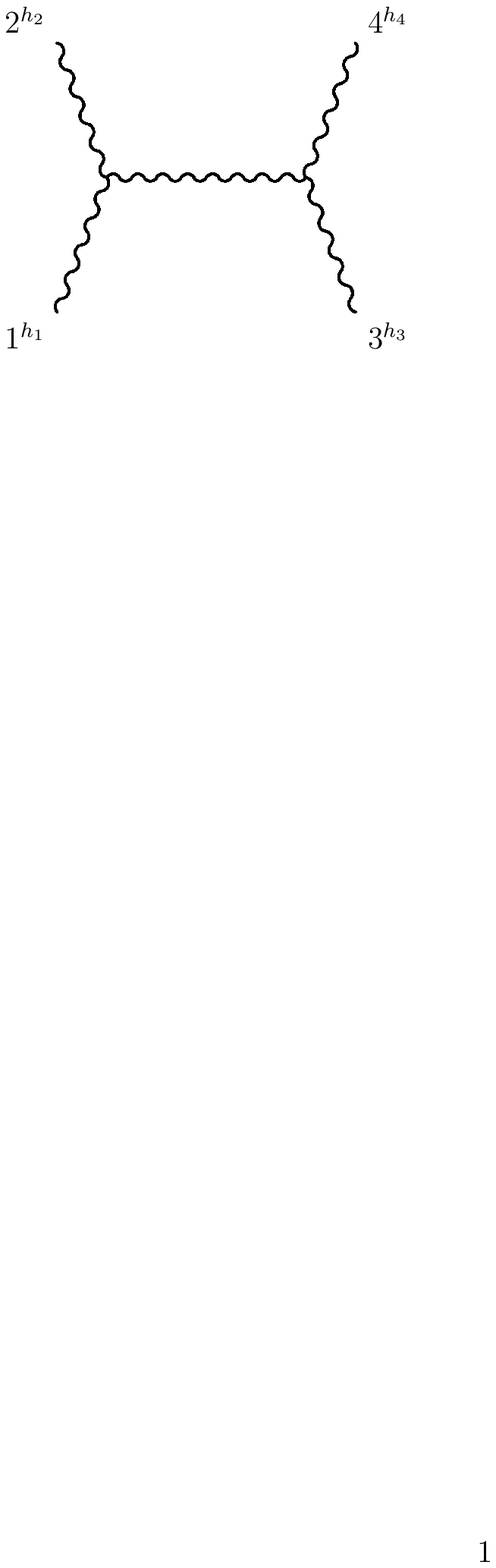}}}}
  \:+\:
  \raisebox{1.9cm}{\scalebox{.85}{{\includegraphics*[104pt,725pt][320pt,607pt]{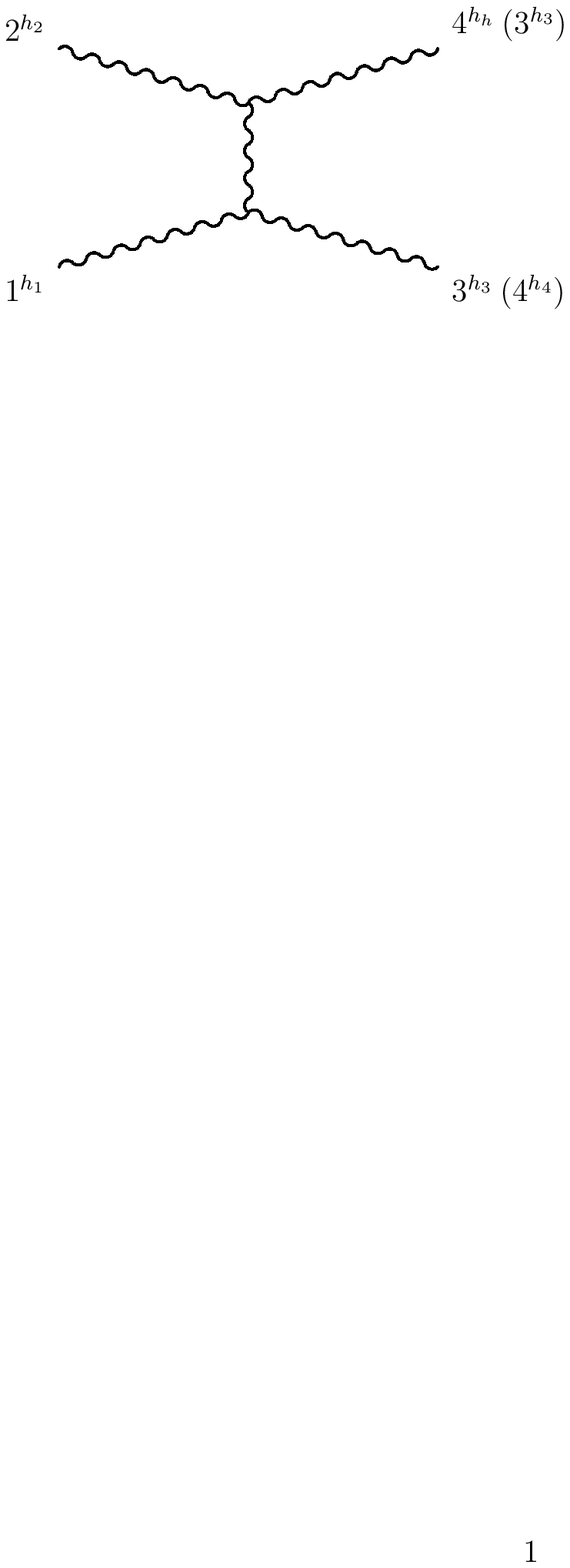}}}}
  \:+\:
  \raisebox{2.15cm}{\scalebox{.85}{{\includegraphics*[116pt,733pt][241pt,598pt]{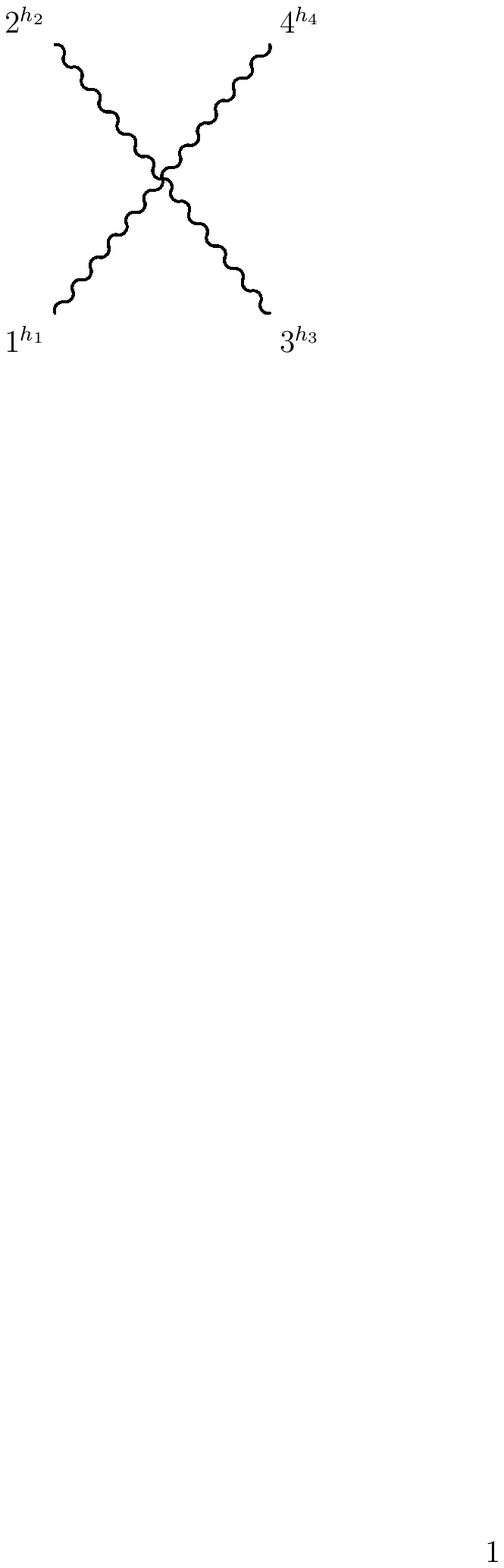}}}}
\]
\vspace{1cm}
\caption{The three different kinds of Feynman diagrams which exhibit different behavior as 
         $z\rightarrow\infty$. They correspond to the ${\bf s}$-channel, 
         ${\bf t}\,({\bf u})$-channel and the four-particle coupling respectively.
         \label{fig:Feynman_diagrams}}
\end{figure}

Feynman diagrams fall into three different categories corresponding to
different behaviors at infinity. Representatives of each type are
shown in figure~~\ref{fig:Feynman_diagrams}. The first kind corresponds to the $(1,2)$-channel
(${\sf s}$-channel). The second corresponds to either the
$(1,3)$-channel (${\sf u}$-channel) or the $(1,4)$-channel (${\sf
t}$-channel). Finally, the third kind is the four-particle coupling.

Under the deformation $\lambda^{(1)}(z) =
\lambda^{(1)}+z\lambda^{(2)}$ and $\tilde\lambda^{(2)}(z) =
\tilde\lambda^{(2)}-z\tilde\lambda^{(1)}$, polarization tensors give
contributions that go as $z^{-s_1}$ and $z^{-s_2}$ respectively in
the case of integer spin and like $z^{-s_1+1/2}$ and $z^{-s_2+1/2}$
in the case of half-integer spin. Recall that we chose particle 1 to
have positive helicity while particle 2 to have negative helicity. Had 
we chosen the opposite helicities, polarization tensors would have
given positive powers of $z$ at infinity. For simplicity, let us
restrict the rest of the discussion in this section to integer spin
particles.

For the first kind of diagrams, only a single three-particle vertex
is $z$ dependent and gives $z^{L_3}$. Combining the contributions we
find $z^{L_3-s_1-s_2}$. Therefore, we need $s_1+s_2>L_3$.

For the second kind of diagrams, two three-particle vertices
contribute giving $z^{L_3+L_3'}$. This time a propagator also
contributes with $z^{-1}$. Combining the contributions we get
$z^{L_3+L_3'-s_1-s_2-1}$. Therefore we need $s_1+s_2>L_3+L_3'-1$.

Finally, for the third kind of diagrams, only the four-particle
vertex contributes giving $z^{L_4}$. Combining the contributions we
find $z^{L_4-s_1-s_2}$. Therefore we need $s_1+s_2>L_4$.

Summarizing, a four-particle amplitude is constructible, i.e.,
$M_4^{(1,2)}(z)$ vanishes as $z\to \infty$ if $s_1+s_2>L_3$,
$s_1+s_2>L_3+L_3'-1$ and $s_1+s_2>L_4$. It is important to mention
that these are sufficient conditions but not necessary. Recall that
we are interested in the behavior of the whole amplitude and not on
that of individual diagrams. Sometimes it is possible that 
the sum of Feynman diagrams vanishes at infinity even though individual ù
diagrams do not.
Also possible is that since our analysis does not take into account
the precise structure of interaction vertices, there might be
cancellations within the same diagram. In other words, our Feynman
diagram analysis only provides an upper bound on the behavior at
infinity.

Let us go back to the example in the previous section. There
$s_1=s_2=s$, $L_3=L_3'=s$. Note that $s_1+s_2>L_3$ implies $s>0$, as
mentioned at the beginning of this section. The second condition is
empty and the third implies that $L_4<2s$. Thus, our conclusions in
the example are valid only if $s>0$ and four-particle interactions
have at most $2s-1$ derivatives. Note that for $s=1$ this excludes
$(F^2)^2$ terms and for $s=2$ this excludes $R^2$ terms. We will
comment on possible ways to make these theories constructible in
section VII.

\subsection{Physical vs. Spurious Poles}

There is an apparent contradiction when in section IV we used that
the only poles of $M_4(z)$ come from propagators and when earlier in
this section we used that polarization tensors behave as $z^{-s}$.

The resolution to this puzzle is very simple yet amusing. Recall
that polarization tensors are defined only up to the choice of a
reference spinor $\mu$ or $\tilde\mu$ of positive or negative
chirality depending on the helicity of the particle. 
The $z$-dependence in polarization tensors comes from the factors 
in the denominator of the form $\langle \lambda(z), \mu\rangle^s$ or
$[\tilde\lambda(z),\tilde\mu]^s$. The deformed spinors are given by
$\lambda(z) = \lambda + z\lambda'$ (or
$\tilde\lambda(z)=\tilde\lambda + z\tilde\lambda'$) where $\lambda'$
(or $\tilde\lambda'$) are the spinors of a different particle. Now
we see that if $\mu$ is not proportional to $\lambda'$ then
individual Feynman diagrams go to zero as $z$ becomes large due to
the $z$ dependence in the polarization tensors. In the same way,
individual Feynman diagrams possess more poles than just those
coming from propagators. Now let us choose $\mu$ proportional to
$\lambda'$. Then the $z$ dependence in polarization tensors
disappears. We then find that individual Feynman diagrams do not
vanish as $z$ becomes large but they show only poles at the
propagators. Recall that we are not interested in individual Feynman
diagrams, but rather in the full amplitude, which is independent of the
choice of reference spinor. Therefore, since $M_4(z)$ vanishes for 
large $z$ for some choice of reference spinors 
it must also do so for any other choice. This means that the pole at
infinity is spurious. Similarly, poles coming from polarization
tensors are spurious as well.

\section{More Examples}

In this section we give more examples of how the four-particle test
can be used to constrain many theories. In previous sections we
studied theories of a single particle of integer spin $s$ and found
that only $s=2$ admits self-interactions. Here we allow for several
particles of the same spin. In this section we consider the
coupling of a particle of spin $s$ and one of spin $2$. The spin $s$
can be integer or half-integer.

\subsection{Several Particles Of Same Integer Spin}

Consider theories of several particles of the same integer spin $s$.
The idea is to see whether allowing for several particles relaxes
the constraint found in section IV.B.2 that sets $s=2$.

We are interested in four-particle amplitudes where each particle
carries an extra quantum number. We can call it a color label. The
data for each particle is thus
$\{\lambda^{(i)},\tilde\lambda^{(i)},h_i,a_i\}$. As discussed in
section III.B, the most general three-particle amplitudes possess
coupling constants that can depend on the color of the particles.
Here we drop the superscripts $H$ and $A$ in order to avoid
cluttering the equations and define $\kappa_{a_1a_2a_3} =
\kappa_{1-s}f_{a_1a_2a_3}$ where $f_{a_1a_2a_3}$ are dimensionless
factors. The subscript $(1-s)$ is the dimension of the coupling
constant.

Repeating the calculation that led to (\ref{seca}) but this time
keeping in mind that we have to sum not only over the helicity of
the internal particle but also over all possible colors, we find
\begin{equation}
\label{kiko} M^{(1,2)}_4(0) = \kappa_{1-s}^2\sum_{a_I} f_{a_1a_4
a_I}f_{a_I a_3a_2} {\cal A} + \kappa_{1-s}^2\sum_{a_I}f_{a_1a_3
a_I}f_{a_I a_4a_2}{\cal B},
\end{equation}
while
\begin{equation}
\label{kika} M^{(1,4)}_4(0) = \kappa_{1-s}^2\sum_{a_I} f_{a_1a_2
a_I}f_{a_I a_3a_4} {\cal C} + \kappa_{1-s}^2\sum_{a_I} f_{a_1a_3
a_I}f_{a_I a_2a_4}{\cal D}
\end{equation}
with
\begin{equation}
\begin{split}
& {\cal A} = \frac{\langle 2,4 \rangle^4}{\langle 1,2 \rangle\langle
2,3 \rangle\langle 3,4 \rangle\langle 4,1
\rangle}\left(\frac{\langle 2,4 \rangle^3 [1,3]}{\langle
1,2\rangle\langle 3,4\rangle}\right)^{s-1},\quad {\cal B} =
\frac{\langle 2,4 \rangle^3}{\langle 1,2 \rangle\langle 4,3
\rangle\langle 3,1 \rangle}\left(\frac{\langle 2,4 \rangle^3
[1,3]}{\langle
1,2\rangle\langle 3,4\rangle}\right)^{s-1}, \\
& {\cal C} =\frac{\langle 2,4 \rangle^4}{\langle 1,2 \rangle\langle
2,3 \rangle\langle 3,4 \rangle\langle 4,1
\rangle}\left(\frac{\langle 2,4 \rangle^3 [1,3]}{\langle
1,4\rangle\langle 2,3\rangle}\right)^{s-1}, \quad {\cal D} =
\frac{\langle 2,4 \rangle^3}{\langle 1,3 \rangle\langle 3,2
\rangle\langle 4,1 \rangle}\left(\frac{\langle 2,4 \rangle^3
[1,3]}{\langle 1,4\rangle\langle 2,3\rangle}\right)^{s-1}.
\end{split}
\end{equation}

In order to understand why we have chosen to factor out the pieces
that survive when $s=1$ let us study this case in detail.

\subsubsection{Spin 1}

Before setting $s=1$ it is important to recall that three-particle
amplitudes for any odd integer spin did not have the correct
symmetry structure under the exchange of particle labels. At the end
of section III, we concluded that if no other labels were introduced
then the three-particle couplings had to vanish. Now we have
theories with a color label. In this case, it is easy to check that
in order to ensure the correct symmetry properties we must require
$f_{a_1a_2a_3}$ to be completely antisymmetric in its indices.

Let us now set $s=1$. The four-particle test requires
$M^{(1,2)}_4(0) - M^{(1,4)}_4(0) = 0$. First note that the factor in
front of ${\cal B}$ and ${\cal D}$ are equal up to a sign (due to
the antisymmetric property of $f$). Therefore they can be combined
and simplified to give
\begin{equation}
\label{jimi} \sum_{a_I}f_{a_1a_3 a_I}f_{a_I a_4a_2}\left({\cal
B}+{\cal D}\right) =-\sum_{a_I}f_{a_1a_3 a_I}f_{a_I a_4a_2}\left(
\frac{\langle 2,4 \rangle^4}{\langle 1,2 \rangle\langle 2,3
\rangle\langle 3,4 \rangle\langle 4,1 \rangle} \right)
\end{equation}
where the right hand side was obtained by a simple application of
the identity $\langle 1,2 \rangle\langle 3,4 \rangle+\langle 1,4
\rangle\langle 2,3 \rangle=\langle 1,3 \rangle\langle 2,4 \rangle$
which follows from the fact that spinors are elements of a
two-dimensional vector space\footnote{Readers familiar with
color-ordered amplitudes possibly have recognized (\ref{jimi}) as
the $U(1)$ decoupling identity, i.e.,
$A(1,2,3,4)+A(2,1,3,4)+A(2,3,1,4) = 0$.}.

Note that the right hand side of (\ref{jimi}) can nicely be combined
with the other terms to give rise to the following condition
\begin{equation}
\sum_{a_I} f_{a_1a_4 a_I}f_{a_I a_3a_2} + \sum_{a_I}f_{a_1a_3
a_I}f_{a_I a_4a_2} + \sum_{a_I} f_{a_1a_2 a_I}f_{a_I a_3a_4} = 0.
\end{equation}
This condition is nothing but the Jacobi identity! Therefore, we
have found that the four-particle test implies that a theory of
several spin $1$ particles can be non-trivial only if the
dimensionless coupling constants $f_{a_1a_2a_3}$ are the structure
constants of a Lie algebra.

\subsubsection{Spin 2}

After the success with spin $1$ particles, the natural question is
to ask whether a similar structure is possible for spin $2$. Once
again, before setting $s=2$ let us mention that like in the case of
odd integer spin particles, the requirement of having the correct
symmetry properties under the exchanges of labels implies that the
dimensionless structure constants, $f_{a_1a_2a_3}$, must be
completely symmetric for even integer spin particles.

Imposing the four-particle test using (\ref{kiko}) and (\ref{kika})
we find that the most general solution requires
\begin{equation}
\label{lasi} \sum_{a_I} f_{a_1a_4 a_I}f_{a_I a_3a_2} =
\sum_{a_I}f_{a_1a_3 a_I}f_{a_I a_4a_2}
\end{equation}
which due to the symmetry properties of $f_{abc}$ implies that all
the other products of structure constants are equal and they factor
out of (\ref{kiko}) and (\ref{kika}) leaving behind the amplitudes
for a single spin 2 particle which we know satisfy the four-particle
test.

Note that (\ref{lasi}) implies that the algebra defined by
\begin{equation}
{\cal E}_a\star {\cal E}_b = f_{abc}~{\cal E}_c
\end{equation}
must be commutative and associative. It turns out that those
algebras are reducible and the theory reduces to that of several
non-interacting massless spin 2 particles. This proves that it is
not possible to define a non-abelian generalization of a theory of
spin 2 particles that is constructible\footnote{We thank L. Freidel
for useful discussions about this point.}. The same conclusion was
proven by using BRST methods in \cite{Boulanger:2000rq}.

Finally, let us mention that for $s>2$ there is no non-trivial
way of satisfying the four-particle test.

\subsection{Coupling Of A Spin s Particle To A Spin 2 Particle}

Our final example of the use of the four-particle test is to
theories of a single spin $s$ particle $(\Psi)$ and a spin 2
particle $(G)$. Here we assume that the spin 2 particle only has
cubic couplings of the form $(++-)$ and $(--+)$. This means that we
are dealing with a graviton. Let the coupling constant of three
gravitons be $\kappa$ while that of a graviton to two $\Psi$'s be
$\kappa'$. Assume that the graviton coupling preserves the helicity
of the $\Psi$ particle. This implies that $\kappa$ and $\kappa'$
have the same dimensions. Also assume that there no any cubic
coupling of $\Psi$'s\footnote{This last condition is not essential
since such a coupling would have dimension different from that of
$\kappa$ and $\kappa'$ and hence it would have to satisfy the
four-particle test independently.}.

We need to analyze two different 4 particle amplitudes:
$M_4(G_1,G_2,\Psi_3,\Psi_4)$ and $M_4(\Psi_1,\Psi_2,\Psi_3,\Psi_4)$.

Consider first $M_4(\Psi_1^{-},\Psi_2^{+},\Psi_3^{-},\Psi_4^{+})$
under a BCFW deformation. A Feynman diagram analysis shows that the
theory is constructible, i.e., the deformed amplitude vanishes at
infinity, for $s>1$. This implies that the following discussion
applies only to particles $\Psi$'s with spin higher than 1.

Let us consider the four-particle test. We choose to deform
$(1^{-},2^{+})$ and $(1^{-},4^{+})$:
\begin{equation}\label{test1}
\begin{split}
M_4^{(1,2)}\:=&\:(\kappa')^{2}
 \frac{\langle1,4\rangle}{[1,4]}
 \frac{[2,4]^{4s}}{[1,2]^{2s-2}[2,3]^{2}[3,4]^{2s-2}}\\
M_4^{(1,4)}\:=&\:(\kappa')^{2}
 \frac{\langle1,2\rangle}{[1,2]}
 \frac{[2,4]^{4s}}{[1,4]^{2s-2}[3,4]^{2}[2,3]^{2s-2}}.
\end{split}
\end{equation}
Notice that $M_4^{(1,4)}$ is obtained from $M_4^{(1,2)}$ by
exchanging $2$ and $4$. Taking the ratio of the quantities in
(\ref{test1}) leads to:
\begin{equation}\label{ratio1}
\frac{M_4^{(1,2)}}{M_4^{(1,4)}}\:=\:
 \left(\frac{\sf t}{\sf s}\right)^{2s-3},
\end{equation}
where ${\sf s} = P_{12}^{2}$ and ${\sf t}=P_{14}^{2}$. This ratio is
equal to one only if $s=3/2$. Thus, the only particle with spin
higher than $1$ which can couple to a graviton, giving a
constructible theory, has the same spin as a gravitino in ${\cal
N}=1$ supergravity.

At this point the couplings $\kappa$ and $\kappa'$ are independent
and it is not possible to conclude that the theory is linearized
supergravity. Quite nicely, the next amplitude constrains the
couplings.

Consider the four-particle test on the amplitude
$M_4(G_1,G_2,\Psi_3,\Psi_4)$. Again we choose to deform $(1,2)$ and
$(1,4)$:
\begin{equation}\label{test2}
\begin{split}
M_{4}^{(1,2)}\:=&\:-(\kappa')^{2}
 \frac{\langle 1,3\rangle^{2}[2,4]^{2s+2}}{[1,2]^{2}[3,4]^{2}[2,3]^{2s-4}}
 \frac{\sf s}{\sf t u}\\
M_{4}^{(1,4)}\:=&\:\kappa' \frac{\langle
1,3\rangle^{2}[2,4]^{2s+2}}{[1,4]^{2}[2,3]^{2s-2}}
 \left(\frac{\kappa}{\sf s}+\frac{\kappa'}{\sf
 u}\right)
\end{split}
\end{equation}
where ${\sf u}=P_{13}^2$.

Taking their ratio and setting $s=3/2$, we get
\begin{equation}\label{ratio2}
\frac{M_{4}^{(1,4)}}{M_{4}^{(1,2)}}\:=\: 1 -\frac{\sf u}{\sf t}
 \left(\frac{\kappa}{\kappa'} - 1\right).
\end{equation}

Requiring the right hand side to be equal to one implies that
$\kappa'=\kappa$. This means that this theory is unique and turns
out to agree with linearized ${\cal N}=1$ supergravity.

An interesting observation is that the local supersymmetry of this
theory arises as an accidental symmetry. The only symmetry we used
in our derivation was under the Poincar{\' e} group; not even global
supersymmetry was assumed. It has been known for a long time
\cite{Grisaru:1976vm} that if one imposes global supersymmetry, then
${\cal N}=1$ supergravity is the unique theory of spin $2$ and spin
$3/2$ massless particles. The uniqueness of ${\cal N}=1$ supergravity
was successively \cite{Deser:1979zb} derived from the non-interactive form
by using gauge invariances. More recently and by using cohomological
BRST methods, the assumption of global supersymmetry was dropped
\cite{Boulanger:2001wq}.

Finally, let us stress that this analysis does not apply to the
coupling of particles with spin $s\le 1$ since the deformed
amplitude under the BFCW deformation does not vanish at infinity.
This simply means that we need to implement our procedure in a
different way. We discuss this briefly in the next section as well
as in the appendix.

\section{Conclusions And Future Directions}

Starting from the very basic assumptions of Poincar{\'e} invariance
and factorization of the S-matrix, we have derived powerful
consistency requirements that constructible theories must satisfy.
We also found that many constructible theories satisfy the
conditions only if the S-matrix is trivial. Non-trivial S-matrices
seem to be rare.

The consistency conditions we found came from studying theories
where four-particle scattering amplitudes can be constructed out of
three-particle ones via the BCFW construction. While failing to
satisfy the four-particle constraint non-trivially means that the
theory should have a trivial S-matrix, passing the test does not
necessarily imply that the interacting theory exists. Once the
four-particle test is satisfied one should check the five- and
higher-particle amplitudes. A theory where all $n$-particle amplitudes
can be determined from the three-particle ones is called fully
constructible.

It is interesting to note that Yang-Mills \cite{Britto:2005fq} and
General Relativity \cite{Benincasa:2007qj} are fully constructible.
This means that the theories are unique in that once the
three-particle amplitudes are chosen (where the only ambiguity is in
the value of the coupling constants) then the whole tree-level
S-matrix is determined. In the case of General Relativity it turns
out that general covariance emerges from Poincar{\'e} symmetry. In
the case of Yang-Mills, the structure of Lie algebras, i.e.,
antisymmetric structure constants that satisfy the Jacobi identity,
also emerges from Poincar{\'e} symmetry. In both cases, the only
non-zero coupling constants of three-particle amplitudes were chosen
to be those of $M_3(++-)$ and $M_3(--+)$. It is important to mention
that our analysis does not discard the possibility of theories with
three-particle amplitudes of the form $M_3(---)$ and $M_3(+++)$.
Dimensional analysis shows that these theories are non-constructible
due to the high power of momenta in the cubic vertex. For example,
if $s=2$ one finds six derivatives. Indeed, for spin $2$, Wald
\cite{Wald:1986bj} found consistent classical field theories that
propagate only massless spin $2$ fields and which are not linearized
General Relativity. Those theories do not possess general covariance
and the simplest of them possesses cubic couplings with six
derivative interactions. In this class of theories might be the spin 3 self-interaction,
which seems to be possible from \cite{Damour:1987fp}, as well as the 
recent proposal for spin 2 and spin 3 interaction of
\cite{Boulanger:2006gr}.

There are some natural questions for the future. One of them is to
ask what the corresponding statements are if one replaces
Poincar{\'e} symmetry by some other group. In particular, it is
known that interactions of higher spins are possible in anti-de
Sitter space (see \cite{Bekaert:2005vh} and references therein). It
would be interesting to reproduce such results from an S-matrix
viewpoint.

The constraints we obtained in this paper only concern the pole
structure of the S-matrix. It is natural to expect that branch cuts
might lead to more constraints. In field theory one is very
familiar with this phenomenon; some theories that are classically
well defined become anomalous at loop level. It would be very
interesting to find out whether the approach presented in this paper
can lead to constraints analogous to anomalies. Speculating even
more, one could imagine that since three-particle amplitudes are
determined exactly, even non-perturbatively, then it might be
possible to find constraints that are only visible outside
perturbation theory.

A well known way to handle quantum corrections is supersymmetry. A
natural generalization of the results of this paper is to replace
Poincar{\'e} symmetry by super Poincar{\'e} and then explore
consistency conditions for theories involving different
supermultiplets.

All of these generalizations, if possible, will only be valid for the
set of constructible theories. In order to increase the power of
these constraints one has to find ways of relaxing the condition of
constructibility. Two possibilities are worth mentioning.

The first approach is to compose several BCFW deformations
\cite{Bern:2005hs} so that more polarization tensors
vanish at infinity and make the amplitude constructible. This
procedure works in many cases but it is not very useful for four
particles since deforming three particles means that one has to sum
over all channels at once and the four-particle constraint is
guaranteed to be satisfied. One can however go to five and more
particles and then there will be non-trivial constraints.

Some peculiar cases can arise because, as it was stressed in section
V, the behavior at infinity obtained by a Feynman diagram analysis
is only an upper bound. It turns out in many examples that a Feynman
diagram analysis shows a non-zero behavior at infinity under a
single BCFW deformation and a vanishing behavior under a composition
of BCFW deformations. Using the composition, one computes the
amplitude which naturally comes out in a very compact form. When one
takes this new compact, but equivalent, form of the amplitude and
looks again at the behavior under a single BCFW deformation, one
finds that it does go to zero at infinity! This shows that there are
cancellations that are not manifest from Feynman diagrams. It would
be very interesting if there was a simple and systematic way of
improving the Feynman diagram analysis so that it will produce
tighter upper bounds. It would be even more interesting to find a way
of carrying out the analysis only in terms of the S-matrix.

The second possibility is to introduce auxiliary massive fields such
that quartic vertices with too many derivatives arise as effective
couplings once the auxiliary field is integrated out. Propagators of
the auxiliary field create poles in $z$ whose location is
proportional to the mass of the auxiliary field. The theory is then
constructible, in the sense that no poles are located at infinity.
Once the amplitudes are obtained one can take the mass of the
auxiliary field to infinity and then recover the original theory.
This gives a nice interpretation to the physics at infinity of some
non-constructible theories: {\it the presence of poles at infinity
implies that the theory is an effective theory where some massive
particles have been integrated out}. The simplest example is 
a theory of a massless scalar $s=0$. Recall that one condition for a
theory to be constructible is that the quartic interaction has to
have $l< 2s$ derivatives. In the case at hand, with $s=0$, this
means that the quartic interaction must be absent. Therefore, a
scalar theory with a $\lambda\phi^4$ interaction is not
constructible. In the appendix, we show that this theory can be made
constructible by introducing an auxiliary field (and deforming three
particles).

A necessary ingredient to carry out the program of auxiliary fields
is to find three-particle amplitudes where one or more particles are
massive. More generally, it will be interesting to extend our
methods for general massive representations of the Poincar{\'e}
group. A good reason to believe that this might be possible is the
analysis of \cite{Badger:2005zh} where amplitudes of massive scalars
and gluons were constructed using a suitable modification of BCFW
deformations. In the case of massive particles of higher spins one
might try to generate a mass term using the Higgs mechanism.

Finally, there are two more directions that, in our view, deserve
further study. The first is the extension to theories in higher or
less number of dimensions, including theories in ten dimensions. The
second is to carry out a systematic search for theories where
several three-particle amplitudes might have coupling constants with
different dimensions but that when multiplied to produce
four-particle amplitudes produce accidental degeneracies. Such
degeneracies might lead to new consistent non-trivial theories which
we might call {\it exceptional theories}.

\begin{acknowledgments}

The authors would like to thank E. Buchbinder, B. Dittrich, L.
Freidel, X. Liu and S. Speziale for useful discussion. PB would like
to thank Perimeter Institute for hospitality during a visit where
part of this research was done. The authors are also grateful
to Natasha Kirby for reading the manuscript. The research of FC at 
Perimeter Institute for Theoretical Physics is supported in part by the
Government of Canada through NSERC and by the Province of Ontario
through MRI.

\end{acknowledgments}

\begin{appendix}

\section{Relaxing Constructibility: Auxiliary Fields}

Our proposal for studying arbitrary spin theories is very general,
but it suffers from the fact that some interesting theories are not
constructible. In section VII, we mentioned several ways of trying
to extend the range of applicability of our technique. One of them
was the introduction of auxiliary fields. In this appendix we
illustrate the idea by showing how the $\lambda \phi^4$ theory,
which is not constructible (even under compositions of BCFW
deformations), can be thought of as the effective theory of a
constructible theory which contains a massive field. The
constructibility here is under a composition of two BCFW
deformations.

The failure to be constructible of the four-particle amplitude in
the $\lambda \phi^4$ theory is understood as a consequence of
sending the mass of the heavy auxiliary field to infinity.

Let us start with a massless scalar with a $\lambda\phi^{4}$
interaction:
\begin{equation}\label{phi4}
\mathcal{L}(\phi)\:=
  \frac{1}{2}\left(\partial_{\mu}\phi\right)\left(\partial^{\mu}\phi\right)
  -\frac{\lambda}{4!}\phi^{4}.
\end{equation}

We can remove the quartic coupling by introducing a massive
auxiliary field $\chi$:
\begin{equation}\label{chiphi2}
\mathcal{L}(\phi,\chi)\:=
  \frac{1}{2}\left(\partial_{\mu}\phi\right)\left(\partial^{\mu}\phi\right)+
  \frac{1}{2}\left(\partial_{\mu}\chi\right)\left(\partial^{\mu}\chi\right)
  -\frac{1}{2}m_{\chi}^{2}\chi^{2}
  -g\chi\phi^{2}.
\end{equation}
It is straightforward to check that (\ref{phi4}) can be obtained
from (\ref{chiphi2}) by integrating out the field $\chi$ taking the
limit of large $g$ and large $m_{\chi}$, and by keeping
$g^{2}/2m_{\chi}^{2}\equiv \lambda/4!$ finite.

The theory (\ref{chiphi2}) now has only cubic interactions. Since
massless scalar fields do not possess polarization tensors that can
be made to vanish at infinity, the theory with only cubic
interactions is still not constructible under a BCFW deformation of
two particles. This problem is resolved by applying a composition
and deforming three particles.

Another problem one has to deal with is that the new vertex in
(\ref{chiphi2}) involves a massive scalar. This implies that the
analysis of section III is not readily applicable. However, in this
specific case, the three particle amplitude is simply given by the
coupling constant $g$.

Since we are interested in the scattering of the massless scalars
represented by the field $\phi$, we consider only amplitudes where
$\chi$ appears as an internal particle. This means that an internal
propagator takes the form
\begin{equation}\label{intprop}
\frac{1}{P^{2}-m_{\chi}^{2}}.
\end{equation}
Let $M_{4}(\phi_{1},\phi_{2},\phi_{3},\phi_{4})$ be the four
particle amplitude of interest. From Feynman diagrams, it is easy to
see that it is given by
\begin{equation}\label{4phiampl}
M_{4}(\phi_{1},\phi_{2},\phi_{3},\phi_{4})\:=\:
 \sum_{j=2}^{4}\frac{g^{2}}{P_{1j}^{2}-m_{\chi}^{2}},
\end{equation}
where $P_{1j}=p^{(1)}+p^{(j)}$. Already from (\ref{4phiampl}), one
can see that the correct limit leads to the four point vertex of the
original theory:
\begin{equation}\label{4philim}
\frac{g^{2}}{P_{1j}^{2}-m_{\chi}^{2}}\:\rightarrow\:
 -\frac{g^{2}}{m_{\chi}^{2}}\:\sim\:\lambda.
\end{equation}

Let us apply a three-particle deformation:
\begin{equation}\label{4phi3def}
\begin{split}
\tilde{\lambda}^{(1)}(z)&=\tilde{\lambda}^{(1)}-
 z\left(\frac{[1,3]}{[2,3]}\tilde{\lambda}^{(2)}
       +\frac{[1,3]}{[3,4]}\tilde{\lambda}^{(4)}\right)\\
\lambda^{(2)}(z)&=\lambda^{(2)}+z\frac{[1,3]}{[2,3]}\lambda^{(1)}\\
\lambda^{(4)}(z)&=\lambda^{(4)}+z\frac{[1,3]}{[3,4]}\lambda^{(1)}.
\end{split}
\end{equation}
A Feynman diagram analysis shows that the deformed amplitude
vanishes at infinity as $z^{-1}$. Taking the ${\sf t}$-channel as an
example, the deformed propagator in this channel is:
\begin{equation}\label{4phidefp}
\frac{1}{P_{14}^{2}(z)-m_{\chi}^{2}}, \quad
P_{14}(z)\:=\:P_{14}-z\frac{[1,3]}{[2,3]}\lambda^{(1)}\tilde{\lambda^{(2)}},
\end{equation}
and its pole is given by
\begin{equation}\label{4phipole}
z_{{\sf u}}\:=\:\frac{[2,3]}{[1,3]}
 \frac{(P_{14}^{2}-m_{\chi}^{2})}{\langle1,4\rangle[2,4]}.
\end{equation}
The momentum $P_{14}$ on-shell becomes:
\begin{equation}\label{4phishell}
P_{14}(z_{{\sf u}})\:=\:P_{14}-
 \frac{(P_{14}^{2}-m_{\chi}^{2})}{\langle1,4\rangle[2,4]}
 \lambda^{(1)}\tilde{\lambda}^{(2)}.
\end{equation}

As stated at the beginning of the appendix, the three-particle
amplitude is just the coupling constant $g$, so it is easy to
reconstruct the result (\ref{4phiampl}) and, as a consequence,
(\ref{4philim}).

\end{appendix}

\end{document}